\documentclass[aps,prd,preprint,tightenlines,superscriptaddress,
amsfonts,amssymb,amsmath,byrevtex,showpacs,nofootinbib,twocolumn]{revtex4-2}
\usepackage[a4paper, margin=0.6in]{geometry}
\usepackage[polish,english]{babel}
\usepackage[applemac]{inputenc}
\usepackage[T1]{fontenc}

\usepackage{latexsym}
\usepackage{graphicx}
\usepackage{geometry}
\usepackage{epstopdf}
\usepackage{subfig}
\usepackage{color}

\usepackage{xcolor}
\usepackage{pdfsync}
\usepackage{fancyhdr}
\usepackage{makeidx}
\usepackage[breaklinks,colorlinks,backref]{hyperref}
\hypersetup{
    colorlinks, 
    linktoc=all, 
    linkcolor=red,  
  citecolor=hyptxt,
  urlcolor=blue}
  \hypersetup{
  citebordercolor=,
  filebordercolor=green,
  linkbordercolor=blue
}
\definecolor{hyptxt}{rgb}{0.7, 0.4, 0.9}
\usepackage{bibtopic}
\newcommand{\bea}{\begin{eqnarray}}
\newcommand{\eea}{\end{eqnarray}}

\newcommand{\ket}[1]{|\kern.3ex#1\kern.3ex\rangle}
\newcommand{\bra}[1]{\langle\kern.3ex #1 \kern.3ex|}
\newcommand{\scalar}[2]{\langle\kern.3ex #1 \kern.3ex|\kern.3ex#2\kern.3ex\rangle}
\newcommand{\norm}[1]{\|\kern.3ex#1\kern.3ex \|}
\newcommand{\be}{\begin{equation}}
\newcommand{\ee}{\end{equation}}

\begin{document}

\title{Some remarks on the horizon in the dust cloud collapse}

\author{Koushiki}
\email{koushiki.malda@gmail.com}
\affiliation{International Centre for Space and Cosmology, Ahmedabad University,
Ahmedabad, GUJ 380009, India}

\author{W{\l}odzimierz Piechocki} \email{wlodzimierz.piechocki@ncbj.gov.pl}
\affiliation{Department of Fundamental Research, National Centre for Nuclear
  Research, Pasteura 7, 02-093 Warsaw, Poland}

\author{Grzegorz Plewa} \email{grzegorz.plewa@pse.pl}
\affiliation{National Centre for Energy Analysis, Sienna 73, 00-833
 Warsaw, Poland}

\date{\today}

\begin{abstract}
We examine the existence of an apparent horizon in the collapse of an isolated dust cloud using expansion functions.
Our results indicate that in the region of spacetime far away from the gravitational singularity, the considered system
has a horizon, in which case the singularity is covered. Using this method may have limited applicability in the neighbourhood
of the gravitational singularity, where quantum effects are expected to be essential.
\end{abstract}


\maketitle

\tableofcontents

\section{Introduction}
Within a year of Einstein's equations coming into existence, an exact solution to them were found in Schwarzschild's solution.
And this  was a boon and a bane, since it contained a singularity at its center. Despite this result, singularities, for a long time,
were considered to be the artifacts of stringent symmetric conditions. Following that, Hawking, Geroch and Penrose's singularity
theorems \citep{HE} proved that singularities are fundamental geometric features of all generic spacetimes except the Minkowskian
flat spacetime. However fundamental these features are, they are problematic since they are the edges of the otherwise smooth,
paracompact, Hausdroff spacetime manifolds \cite{Joshibook1} and the maximal Cauchy development of inextendible non-spacelike
geodesics on these manifolds are truncated at these edges. Therefore, spacetime singularities are generic and irremovable features
of general relativity and the biggest problem surrounding these objects are their genericity and visibility, the second
one we discuss in this work.

Schwarzschild's solution and the singularity contained in it is a unique solution for vacuum,  asymptotically flat, static,
spherically symmetric  spacetimes \cite{GosEllis}. This specific solution can also be obtained from the collapse of spherically
symmetric,  pressureless and homogeneous dust cloud \cite{Datt, Oppenheimer1939}. This work also showed that an event
horizon\footnote{It is defined as the boundary of the causal past of future null infinity. So, to obtain the event horizon,
the asymptotic null infinity has to be known for the particular spacetime. Hereby this definition is teleological, not local.}
would form before the formation of this spacelike singularity and so, any causal connection in the direction of the future f
rom this is automatically prohibited: this singularity is a blackhole.

Following the flawless structure of the Schawarzschild blackhole, it was hypothesised by Penrose \cite{CCC, Penrose1979SCC}
that all spacetime singularities should be hidden behind event horizons and no future directed non-spacelike geodesic is allowed to
escape from the neighbourhood of it. However, there is no proof for this theorem:  Cosmic Censorship Conjecture, but it
was assumed to be true overarchingly in classical theory of gravity. It is important to emphasize at this point that the singularity
theorems do not necessitate the existence of trapped surfaces or horizons for the existence of an incomplete non-spacelike geodesic,
which is a necessary and sufficient condition for a singularity to exist. Following this, multiple different physically reasonable
matter fields (of types I and II \cite{HE}) were investigated to resolve the issue of visibility in spacetime singularities and
some counter-intuitive results were obtained. Christodoulou found out that a singularity can be formed from a dust cloud collapse
where there are no horizons forming \cite{Christodoulou:1984mz}. Later works also confirmed these results \cite{Newman:1985gt, Newman:1988cr}.
It was shown that the inhomogeneous, self-similar, marginally bound, pressureless dust, after going through a continual collapse,
produces a singularity where a local or apparent horizon is formed at the same instance when the singularity forms. Investigation
of the null geodesics  showed that, unlike a blackhole, the tangents to the future directed ones were positive
definite \cite{Joshi1} and thus it was structurally different from a blackhole: a  naked singularity. Based on these findings,
a precise definition of a naked singularity was given: if there exists a family of inextendible outgoing non-spacelike geodesics
in a spacetime, whose causal past lie in the neighbourhood of the singularity, then it is at least locally naked. These results
were also obtained for non self-similar cases \cite{Joshi2, Joshi3} as well as non-marginally bound cases \cite{Mosani:2020mro}.
The results regarding an end-state singularity stayed qualitatively similar when the pressure was turned on in the collapsing
cloud \cite{Madhav:2005kg, Goswami:2006ph, Koushiki:2025sce, Koushiki:2025uqr}. However, these results of naked singularity only
regard visibility locally without discussing whether the geodesics are future complete: whether they will reach an observer at
asymptotic distance.

Global visibility becomes important for observational purposes, since locally naked singularities would be more difficult to
locate and gather signals from them using distant probes. There are enough examples of globally naked singularities as well \cite{Joshi:2011zm, Joshi:2013dva, Goswami:2007na, Mosani:2021byw, Mosani:2023vtr, Koushiki:2025dax}. More importantly, these models have found relevance in recent observational studies: the central compact object of the Sgr A* has been stipulated as either one of a black-hole or a naked singularity \cite{EventHorizonTelescope:2022wkp}. Naked singularities have found substantial support from different studies involving simulations, for instance accretion flows \cite{Kluzniak:2024cxm, Saurabh:2023otl, Uniyal:2025hik}, shadows \cite{Dey:2020bgo, Joshi:2020tlq}, precession orbits \cite{Solanki:2021mkt, Dey:2020haf} etc.

Globally naked singularities are important since their existence allows very strong gravity regimes come into causal contact to even asymptotic observers since there is no horizon prohibiting this contact. As quantum gravitational effects are said to become important at such length and energy scales, their direct observation becomes possible if the singularity is globally visible. Quantum approaches have been applied to gravitational collapse scenarios of homogeneous, pressureless, spherically symmetric clouds \cite{Gozdz:2023ddu, Schmitz:2019jct}. One similar approach has been applied to a homogeneous collapsing cloud with non-zero isotropic pressure \cite{Goswami:2005fu}, in which the singularity evaporates on application of the quantum theory. So, the next natural step would be to apply quantum theory to a more complex collapsing scenario: an inhomogeneous, but pressureless collapsing cloud. The reason to explore this direction is to find the fate of gravitational singularities at quantum level. That would also help to construct a satisfactory theory of quantum gravity.

We build the premise of the same in the present work and reinvestigate the features of an inhomogeneous, pressureless, spherically symmetric, marginally bound dust. The aim of this paper is to examine the existence of the horizon and the trapped region of the isolated collapsing cloud of dust using expansion functions, since we are interested in astrophysically relevant naked singularities, where horizons do not at all form. In what follows we set $8\pi G/c^4 =1.$

\section{The LTB spacetime}

The gravitational collapse of a spherically symmetric, inhomogeneous and marginally bound cloud of dust can be described  by the  Lema\^{i}tre-Tolman-Bondi
metric \cite{Joshi1,Joshi3,Singh}
\bea
\label{met}
ds^2 = -d t^2 + (R')^2 \, d r^2 +
\nonumber \\ R^2 (d \theta^2 + \sin^2\theta\,d \phi^2)\, ,
\eea
where $R = R(t,r)$ is the area radius at time $t$ of the shell having radial coordinate  $r$, and where $R' = \partial R/\partial r$.
The energy-momentum tensor $T^{\mu\nu} = \rho (t,r) U^\mu U^\nu$, where $\rho (t,r)$ is the energy density of the cloud and $U^\mu$
is the four velocity.

Inserting the metric \eqref{met} into the Einstein equations leads to
\begin{equation}\label{mass}
\dot{R}^2 = \frac{F}{R}
\end{equation}
and
\begin{equation}\label{rho}
 \rho = \frac{F'}{R^2 R'}\, ,
\end{equation}
where $\dot{R} = \partial R/\partial t$, and where $F = F(r)$ is the mass of the cloud inside the shell of coordinate $r$.

The case of collapse is describe by $R' < 0$, where  $R$ is the physical radius of the collapsing configuration.
The shell index $r$ can be rescalled by demanding $R(0,r) = r$.

Integrating  \eqref{mass} gives
\bea
R^\frac{3}{2}= r^\frac{3}{2}-\frac{3}{2} t\sqrt{F}\, .
\label{Rsol}
\eea
The solution to \eqref{rho} can be found to be
\bea
\label{den} \rho (t,r) = \frac{4 \sqrt{F(r)} F'(r)}{\left(3 t \sqrt{F(r)} -2 r^{3/2} \right)}\nonumber \\ \times \frac{1}{\left(t F'(r)-2
\sqrt{r F(r)}\right)} \, .
\eea
For details, see  App.~\ref{dyn}.

\section{Apparent horizon}

Let $\Sigma$ stands for codimension-two hypersurface embedded in a four-dimensional spacetime. The normal space
to such a surface is spanned by by a pair of null vectors $l^\mu$ and $n^\mu$. Assuming they are cross-normalized,
$l \cdot n = -1$, the induced metric on $\Sigma$ reads \cite{Booth,Abhay,Badri}
\bea\label{ind}
q_{\mu \nu} = g_{\mu \nu} + l_\mu n_\nu + l_\nu n_\mu \, ,
\eea
where $g_{\mu \nu}$ stands for the original metric \eqref{met}. Suppose that $l^\mu$, and $n^\mu$ are future-pointing
vectors. Let $l^\mu$ be an outward pointing to $\Sigma$ while $n^\mu$ points inwards. The corresponding expansions are:
\begin{align}
  \theta_l = q^{\mu \nu} l_{\nu;\mu} \, ,
\\[1ex]
  \theta_n = q^{\mu \nu} n_{\nu;\mu} \, .
\end{align}

The trapped region is defines as a region where both expansions are strictly negative. The apparent horizon  is defined
as a boundary of the trapped region by
\bea
\label{hor} \theta_l = 0, \quad \theta_n < 0 \, .
\eea

To find the position of the horizon, we  restrict ourselves to $t-r$ components and take
\begin{align}
  l^{\mu} = (l^0, l^1, 0, 0)\, ,
  \\[1ex]
  n^{\mu} =  (n^0, n^1, 0, 0).
\end{align}
This is justified by the spherical symmetry, i.e., the fact that the position of the horizon is not affected by the components
$\theta$ and $\phi$. To get the four unknown non-zero components, consider the following conditions
\begin{align}
  \label{init}
  l_\mu l^\mu = n_\mu n^\mu = 0 \,,
  \\[1ex]
  \label{cross}
  l_\mu n^\mu = - 1 \,,
  \\[1ex]
  \label{aux}
  n^1 = - \beta \, l^1 \, ,
\end{align}
supplemented by the requirement $l^0, n^0 > 0$ (future-pointing) and $n^1 \leq 0$, $l^\mu \geq 0$. Future-pointing null geodesics fulfill the requirement
$l_\mu e^\mu >0$, and similarly, past-pointing ones are expressed by $n_\mu e^\mu <0$ with $e^\mu$ being the unit vector in the radial direction.
Eqs.~\eqref{init}--\eqref{cross}
are basic relations for null vectors (the fact they are null and cross-normalized). The condition \eqref{aux} is an auxiliary equation
used for convenient parametrization of vectors pointing inwards and outwards with $\beta  = \beta(t, r) > 0$. Here we require that the
radial components have opposite signs. Solving the system \eqref{init}--\eqref{aux} one finds
\begin{align}
l^0 = \frac{1}{\sqrt{2 \beta}}, \quad l^1 = \frac{1}{\sqrt{2 \beta} R'}\, ,
\\[1ex]
n^0 = \frac{\sqrt{\beta}}{\sqrt{2}}, \quad n^1 = - \frac{\sqrt{\beta}}{\sqrt{2} R'} \, .
\end{align}

Taking this into account, one finds the following non-zero components of the induced metric \eqref{ind}:
\be
q_{\theta \theta} = R^2, \quad q_{\phi \phi} = R^2 \sin^2 \theta.
\ee
Together with the form of $l^\mu$ and $n^\mu$, this leads to
\begin{align}
  \theta_l = \frac{\sqrt{2} (1+\dot{R})}{\sqrt{\beta} R} \, ,
  \label{th l}
\\[1ex]
  \theta_n = \frac{\sqrt{2 \beta} (-1+\dot{R})}{R} \, .
  \label{th n}
\end{align}

It is clear, due to Eqs.~\eqref{hor}, that the apparent horizon corresponds to $\dot{R} = -1$.

Note that the auxiliary function $\beta(t,r)$ does not affect the condition specifying the position of the horizon (assuming that
$\beta \neq 0$), but affects the form of both expansions $\theta_l$ and $\theta_n$.

Supplementing Eqs. \eqref{th l}--\eqref{th n} with the exact solution for dust \eqref{Rsol}, one finds
  \begin{align}
    \label{theta_l}
    \theta_l &= \frac{\sqrt{2} ( \sqrt{R}-\sqrt{F} )}{\sqrt{\beta} R^{3/2}},
    \\[1ex]
    \label{theta_n}
     \theta_n &= -\frac{\sqrt{2 \beta} ( \sqrt{R}+\sqrt{F}  )}{R^{3/2}}.
    \end{align}

Thus,   $\theta_n < 0$ identically. It results from Eq. \eqref{theta_l} that the trapped region is given by
    \be\label{trapped}
  R \leq F \, .
    \ee
The boundary of the trapped region, the apparent horizon, corresponds to the condition  $\theta_l = 0$,
which implies  due to \eqref{theta_l} that we have
\be\label{apphor}
 R = F.
\ee

\section{Matching conditions}

To understand the dynamics, the collapsing LTB cloud has to be stitched to an external space-time.
And these two space-times, stitched at a boundary, then describes the collapsing inhomogeneous dust cloud
surrounded by another static space-time:  the Schwarzschild space-time (due to Birkhoff's theorem).
To do this, we use the Israel-Darmois matching conditions \cite{Poisson}. These conditions require the first fundamental
form (induced metric) and the second fundamental form (extrinsic curvature) to be equal at the hypersurface  separating
two spaces.

For the matching we use the LTB metric inside the most external shell, while
the exterior geometry is given by the Schwarzschild metric.

\subsection{LTB geometry} \label{ltb_part}

The radial coordinate $r$ is fixed at the outermost external shell:
  \be
  \label{rb_eq}
  r = r_b \, .
  \ee
The corresponding hypersurface is defined by
  \be
  \label{LTB hyper}
  R(t,r) = fixed,
  \ee
where $t$, $r$ are LTB coordinates of the metric \eqref{met}. Eq. \eqref{LTB hyper} should be taken at $r = r_b$.

The normal to the hypersurface \eqref{LTB hyper} is
\be
\label{LTB normal}
n_\alpha = - \partial_\alpha R = \{ -\dot{R}, -R', 0, 0 \}.
\ee

Let $y^a$ stand for coordinates on the hupersurface. Because of spherical symmetry, one can chose them as
$\{\lambda, \theta, \phi \}$, where $t$ and $r$ are expected to be a functions of $\lambda$:
$t = t(\lambda)$, $r = r(\lambda)$. Following the notation of \cite{Poisson}, the tangent to the surface reads
\be
e^\alpha{}_a = \frac{\partial x^\alpha}{\partial y^a},
\ee
such that $e{^\alpha_a} n_\alpha = 0$,  which for the normal \eqref{LTB normal} gives
\be
\label{TB rel}
\frac{\partial r}{\partial \lambda} = -\frac{\dot{R}}{R'} \frac{\partial t}{ \partial \lambda}.
\ee
The induced metric is given by the standard formula
\be
G_{a b} = g_{\mu \nu} \frac{\partial x^\mu}{d y^a} \frac{\partial x^\nu}{d y^b} \,,
\ee
where $ g_{\mu \nu}$ is the metric \eqref{met}.
Using the relation \eqref{TB rel} one finds
\bea
G_{a b}^{LTB} d y^a d y^b &=& -\left( 1-\dot{R}^2  \right) \left( \frac{\partial t}{\partial \lambda} \right)^2
d \lambda^2\nonumber\\ &+& R^2 d \Omega^2 \,.
\eea
Without  loosing of generality one can identify $\lambda$ with LTB time $t$. Letting
\be
\lambda = t,
\ee
one gets a simpler form of the induced metric
\be
\label{LTB first fundamental}
G_{a b}^{LTB} d y^a d y^b = -\left( 1-\dot{R}^2  \right) d t^2 + R^2 d \Omega^2 \, .
\ee

The extrinsic curvature \cite{Poisson} reads
\be
K_{a b} = n_{\alpha;\beta}\, e^{\alpha}{}_a e^{\beta}{}_b.
\ee
There are only three non-zero components of the extrinsic curvature:
\begin{align}
  \nonumber
  K^{LTB}_{t t} = - \ddot{R},
  \\[1ex]
  \nonumber
  K^{LTB}_{\theta \theta} = - R(1-\dot{R}^2),
  \\[1ex]
  \label{second fund LTB}
   K^{LTB}_{\phi \phi} = - R(1-\dot{R}^2) \sin^2{\theta}.
\end{align}
%
\subsection{Schwarzschild geometry}

Let $t_s$ and $r_s$ denote the Schwarzschild time and radial coordinates.  The exterior metric reads
  \bea
  \label{Schw}
ds^2 = &-&\left(1-\frac{2M}{r_s}\right) d t_s^2 + \left(1-\frac{2M}{r_s} \right)^{-1}\nonumber\\
&\times & d r_s^2 + r_s^2 d \Omega^2 \,.
\eea
The metric is singular at  $r_s=2 M$, causing significant problems with matching. To avoid them, it is customary
to consider regular coordinates. Below we consider Lemaitre coordinates $\tau, \rho, \theta, \phi$.
These can be found replacing Schwarzschild $t$ and $r$ by $\tau$ and $\rho$ such that:
\begin{align}
  \nonumber
  d \tau = d t + \sqrt{\frac{2 M}{r}} \left( 1-\frac{2 M}{r} \right)^{-1} d r,
  \\[1ex]
  \label{Lemaitre trans}
  d \rho = d t + \sqrt{\frac{r}{2 M}} \left( 1-\frac{2 M}{r} \right)^{-1} d r \,.
\end{align}
Applying transformations \eqref{Lemaitre trans} to the Schwarzschild metric \eqref{Schw} one gets
\be
 d s^2 = - d \tau^2 + \frac{2 M}{r_s} d \rho^2 + r^2_s d \Omega^2,
\ee
where
\be
\label{rs_form}
r_s = (2M)^{1/3} \left[ \frac{3}{2} (\rho-\tau) \right]^{2/3}
\ee
is a function of the new coordinates $\tau$, $\rho$.
Now consider a hupersurface $r_s(\tau, \rho) = const$. This will be an equivalent of the LTB hypersurface  $R(t, r) = const$.
The normal vector reads
\be
n_\alpha = - \partial_\alpha r_s = \{ -\dot{r_s}, -r_s', 0, 0 \},
\ee
where
\be
\label{Lemaitre metric}
\dot{r_s} = \frac{\partial r_s}{\partial \tau}, \quad r_s' = \frac{\partial r_s}{\partial \rho}.
\ee
As in the case of the LTB metric, we install coordinates at a hypersurface $y^a = \{ \lambda, \theta,
\phi \}$ such that $\tau = \tau(\lambda)$, $\rho = \rho(\lambda)$. The condition  $e{^\alpha_a} n_\alpha = 0$ translates into
\be
\label{ne_cond}
\frac{\partial \rho}{\partial \lambda} = - \left( \frac{\partial r_s}{\partial \rho}  \right)^{-1} \frac{\partial r_s}{\partial
\tau} \frac{\partial \tau}{\partial \lambda} \,.
\ee
Using \eqref{ne_cond} one finds the induced metric
\bea
G_{a b}^{Schw} d y^a d y^b = &-& \left( 1-\frac{2 M}{r_s} \left( \frac{\partial r_s}{\partial \rho} \right)^{-2}
\left( \frac{\partial r_s}{\partial \tau} \right)^2   \right)\nonumber \\ &\times & \left( \frac{\partial \tau}{\partial \lambda} \right)^2
d \lambda^2 + r_s^2 d \Omega^2 \,.
\eea
Using Eq. \eqref{rs_form} one gets
\be
\left( \frac{\partial r_s}{\partial \rho} \right)^{-1} \left( \frac{\partial r_s}{\partial \tau} \right) = -1.
\ee
This simplifies the induced metric
\bea
G_{a b}^{Schw} d y^a d y^b = &-& \left( 1-\frac{2 M}{r_s}  \right) \left( \frac{\partial \tau}{\partial \lambda} \right)^2 d \lambda^2 \nonumber \\
&+& r_s^2 d \Omega^2 \,.
\eea
Again, identifying for simplicity
\be
\lambda = \tau,
\ee
one gets
\be
\label{Schw first fundamental}
G_{a b}^{Schw} d y^a d y^b = -\left( 1-\frac{2 M}{r_s}  \right) d \tau^2 + r_s^2 d \Omega^2.
\ee
The extrinsic curvature reads
\begin{align}
  \nonumber
  K^{Schw}_{\tau \tau} = - \ddot{r}_s,
  \\[1ex]
  \nonumber
  K^{Schw}_{\theta \theta} = - r_s \left( \frac{r_s {r_s'}^2}{2 M} - \dot{r_s}^2 \right),
  \\[1ex]
  \label{second fund Schw}
   K^{Schw}_{\phi \phi} = - r_s \left( \frac{r_s {r_s'}^2}{2 M} - \dot{r_s}^2 \right) \sin^2{\theta}.
\end{align}
%
\subsection{Matching}

Comparing the LTB first fundamental form \eqref{LTB first fundamental} with \eqref{Schw first fundamental} gives
\be
\tau = t, \quad r_s = R,
\ee
and
\be
\label{first}
\dot{R}^2 = \frac{2 M}{R}.
\ee
Note that $r_s(t, \rho) = R(t, r)$ implies
\be
\rho = r \, .
\ee
Taking into account Eq.~\eqref{rb_eq} means that $\rho$ is also fixed at the boundary hypersurface, $\rho = r_b$.
However, the original Schwarzschild radial coordinate $r_s$ is not constant.

Due to Eq.~\eqref{mass}, Eq.~\eqref{first} can be alternatively rewritten as
\be
\label{F2M1}
F = 2M.
\ee
Comparing second fundamental forms, i.e. Eq. \eqref{second fund LTB}  and Eq. \eqref{second fund Schw} leads to additional constraint
\be
\label{second}
{R'}^2 = \frac{2 M}{R} \,.
\ee
%
\subsection{Solutions}
From Eq.  \eqref{first} one finds
\be
\label{dot_R}
 \dot{R} = - \sqrt{\frac{2 M}{R}} \,.
 \ee

 The minus sign reflects the effect of gravitational collapse ($R$ decreases with time). Solving Eq. \eqref{dot_R} one finds
 \be
 \label{R0}
 R(t,r_b) = \left( \frac{3}{2} \right)^{2/3} M^{1/3} \left(c(r_b) - \sqrt{2} t \right)^{2/3},
 \ee
 where $c(r_b)$ stands for an arbitrary function taken at $r=r_b$ (as required by the condition \eqref{rb_eq}).
 It is easy to check that Eq. \eqref{R0} implies
 \be
 \label{F2M2}
 \dot{R}^2 R = 2 M
 \ee
 or alternatively
 \be \label{FF2M2}
 F = 2 M.
 \ee
 The condition \eqref{second} gives
 \be
 c'(r_b)^2 = 2,
 \ee
 where we again used the relation \eqref{rb_eq}.

\section{Conclusions}

During the collapse, the amount of matter within the outermost shell $r_b$ is constant,  since $M$ is constant   as shown by Eqs.~\eqref{F2M1} and \eqref{FF2M2},
and reads
\begin{equation}\label{const}
  F(r_b) = 2 M \,.
\end{equation}

The existence of the trapped region and the apparent horizon may occur if we have, owing to Eq.~\eqref{trapped}, the following
\begin{equation}\label{trap2}
  R(t,r) \leq F(r) \, .
\end{equation}
Suppose at the beginning of the star's collapse, we have
\begin{equation}\label{trap3}
  R(t,r) > F (r)
\end{equation}
so that there is no apparent horizon. During the evolution to the gravitational singularity  (see App. \ref{sing}),
the radius $R$ of the collapsing star decreases. This is why Eq.~\eqref{trap3} turns into Eq.~\eqref{trap2}.
Therefore, the considered gravitational system becomes a black hole with the trapped region covered by the apparent horizon.

The above reasoning is justified in the region of a collapsing star far away from the vicinity of a gravitational singularity,
where general relativity is well defined. The situation may be different in the neighbourhood of the singularity, as the full
applicability of general relativity is expected to be limited due to quantum effects. There, using the expansion functions to
derive \eqref{trapped} and \eqref{apphor} may be questionable. This may concern situations in which $F(r_b)$ is so small that
satisfying \eqref{apphor} requires entering the Planck scale.

The results obtained a long time ago \cite{Joshi1,Joshi3,Singh,Joshi:2001xi} on dust cloud collapse suggest the possibility of naked
singularities. The corresponding analyses, however, are  based on the geometry of spacetime in the neighbourhood of the
gravitational singularity. Possible quantum contributions have been  ignored.

Further examination is needed that accounts for quantum effects. This issue will be considered elsewhere.


\appendix

\section{Solution to classical dynamics}
\label{dyn}
Inserting the metric \eqref{met} into the  Einstein equations for dust, one finds the following three independent equations:
\begin{align}
  \label{ein0}
  \frac{\dot{R}^2}{R^2} + \frac{2 \dot{R} \dot{R}'}{R \, R'} = \rho,
  \\[1ex]
  \label{ein1}
  \frac{R'}{R} \left(\dot{R}^2 + 2 R \ddot{R} \right) = 0,
  \\[1ex]
  \label{ein2}
  \frac{R}{R'} \left( \dot{R} \dot{R}' + R' \ddot{R} + R \ddot{R}' \right) = 0,
\end{align}
where $R = R(t,r)$, $\rho = \rho(t,r)$.
Eqs. \eqref{ein1}--\eqref{ein2} can be used to eliminate $\ddot{R}$ and $\ddot{R}'$ derivatives
\begin{align}
  \ddot{R} = - \frac{\dot{R}^2}{2 R},
  \\[1ex]
  \ddot{R}' = \frac{R' \dot{R}^2}{2 R^2} - \frac{\dot{R} \dot{R}' }{R}.
  \end{align}

It can be verified that Eqs.~\eqref{ein1}--\eqref{ein2} are satisfied by $R(t,r)$ that reads
\be
R(t, r) = \left( r^\frac{3}{2}-\frac{3}{2} t\sqrt{F(r)}  \right)^{2/3}.
\ee
On the other hand, from \eqref{ein0} one gets the relation between energy density and $F(r)$
\bea
\label{density} \rho (t,r) = \frac{4 \sqrt{F(r)} F'(r)}{\left(3 t \sqrt{F(r)} -2 r^{3/2} \right)}\nonumber \\
\times \frac{1}{\left(t F'(r)-2 \sqrt{r F(r)}\right)}.
\eea
Hence, the solution to the Einstein equation gives $R$ and $\rho$ to be a functions of the mass $F(r)$.

\section{Curvature invariants}\label{sing}

Curvature invariants are used to measure the geometric curvature of spacetime.
Most common are the Ricci and Riemann invariants.
From the metric \eqref{met} one finds that the Ricci scalar reads
\be
\mathcal{R} :=g_{\mu \nu} R^{\mu \nu}= \frac{2 \dot{R}\dot{R}'}{R R'} + \frac{\dot{R}^2}{R^2} \, .
\ee

The Kretschmann scalar is found to be
\be
\label{Kretsch_final}
\mathcal{K}:= R_{\mu \nu \delta \gamma}R^{\mu \nu \delta \gamma} = 7 \frac{\dot{R}^4}{R^4} - 4 \frac{\dot{R}^3
\dot{R}'}{ R^3 R'} + 12 \frac{\dot{R}^2 \dot{R}'^2}{R^2 R'^2} \, .
\ee


\end{document}